\documentclass[11pt]{article}
\usepackage[final]{acl}

\usepackage{times}
\usepackage{latexsym}
\usepackage{booktabs} 
\usepackage{multirow}
\usepackage{amsmath} 
\usepackage{array} 
\usepackage{amssymb} 
\usepackage{minted}
\usepackage{xcolor}
\usepackage{xcolor}
\usepackage{listings}

\definecolor{LightGray}{gray}{0.9} 

\lstdefinestyle{promptstyle}{
    backgroundcolor=\color{LightGray},
    frame=lines,
    framesep=2mm,
    basicstyle=\linespread{1.2}\footnotesize\ttfamily, % Replicates minted's baselinestretch
    breaklines=true,
    breakatwhitespace=true,
    columns=fullflexible,
    xleftmargin=2mm,
    xrightmargin=2mm,
    aboveskip=1em,
    belowskip=1em,
    showstringspaces=false
}
\usepackage{float}
\usepackage{listings}
\usepackage[utf8]{inputenc}
\usepackage[T1]{fontenc}

\usepackage{microtype}

\usepackage{inconsolata}

\usepackage{graphicx}
\usepackage{tabularx}

\nolinenumbers

\title{HotelQuEST: Balancing Quality and Efficiency in Agentic Search}

\author{Guy Hadad \\
  Ben-Gurion University\\
  \texttt{guyhada@post.bgu.ac.il} \And
  Shadi Iskander \\
  Amazon\\
  \texttt{shadisk@amazon.com} \And
  Oren Kalinsky \\
  Amazon\\
  \texttt{orenk@amazon.com}
  \AND
  Sofia Tolmach \\
  Amazon\\
  \texttt{sofiato@amazon.com} \And
  Ran Levy \\
  Amazon\\
  \texttt{ranlevy@amazon.com} \And
  Haggai Roitman \\
  Amazon\\
  \texttt{hroitman@amazon.com}}
\iffalse\else
\newcommand{\guy}[1]{}
\newcommand{\shadi}[1]{}
\newcommand{\sofia}[1]{}
\newcommand{\yuval}[1]{}
\newcommand{\haggai}[1]{}
\newcommand{\ran}[1]{}
\newcommand{\arnon}[1]{}
\newcommand{\oren}[1]{}
\fi

\def\x{HotelQuEST}
\def\xspace{HotelQuEST }

\begin{document}
\maketitle
\begin{abstract}

Agentic search has emerged as a  promising paradigm for adaptive retrieval systems \mbox{powered} by large language models (LLMs). However, existing benchmarks primarily focus on quality, overlooking efficiency factors that are critical for real-world deployment. Moreover, real-world user queries often contain underspecified preferences, a challenge that remains largely underexplored in current agentic search evaluation. As a result, many agentic search systems remain impractical despite their impressive performance.
In this work, we introduce \textbf{\x}\footnote{Data and code can be found at \url{https://github.com/amazon-science/hotel-quest-benchmark}.}, a benchmark comprising 214 hotel search queries that range from simple factual requests to complex queries, enabling evaluation across the full spectrum of query difficulty. We further address the challenge of evaluating underspecified user preferences by collecting clarifications that make annotators' implicit preferences explicit for evaluation.
We find that LLM-based agents achieve higher accuracy than traditional retrievers, but at substantially higher costs due to redundant tool calls and suboptimal routing that fails to match query complexity to model capability. Our analysis exposes inefficiencies in current agentic search systems and demonstrates substantial potential for cost-aware optimization.

\end{abstract}

\section{Introduction}

\begin{figure}[t]
\centering
\includegraphics[width=1.0\linewidth, height=6cm]{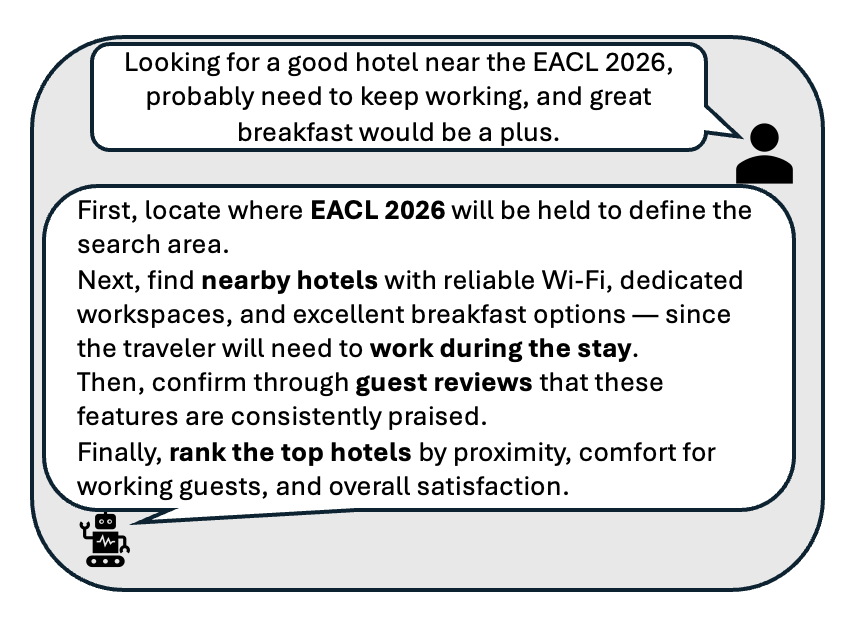}
\caption{
Illustration of a task from our benchmark.}
\label{fig:agent-plan}
\vspace{-0.2cm}
\end{figure}

LLMs have enabled a new generation of autonomous agents that can navigate websites, operate tools, and assist in complex tasks \cite{wang2024survey, zheng2024gpt, xie2024travelplanner, chen2024scienceagentbench}. %Among the most promising applications is agentic search: systems that can iteratively reason, retrieve information, and synthesize answers to satisfy natural language queries \cite{zhang2025web, li2025webthinker, han2025deep}. However, most real-world search workloads vary; systems must handle a high volume of simple queries efficiently while also supporting complex, multi-hop requests that require deeper reasoning \cite{suri2024use}.  
A key emerging application is agentic search, systems that iteratively reason, retrieve information, and synthesize answers to natural-language queries~\cite{zhang2025web, li2025webthinker, han2025deep}. In practice, search workloads vary widely: systems must process large volumes of simple queries efficiently while still handling complex, multi-hop questions that demand deeper reasoning~\cite{suri2024use}.

Existing benchmarks for agentic search focus primarily on answer quality \cite{gou2025mind2web, du2025deepresearch}, neglecting two critical dimensions for practical deployment: (i)~efficiency constraints (latency, cost) that determine practical deployability  \cite{kapoor2024ai}, and (ii)~underspecified user preferences that challenge standard relevance notions~\cite{xi2025infodeepseek, mialon2023gaia}. For instance, ``dog-friendly'' could mean pets are allowed for a fee, allowed freely, or only in certain areas \cite{choi2025bloomintent}. %These gaps make it difficult to determine when agents allocate resources appropriately versus when they over-invest computation for marginal gains.
These gaps make it hard to judge whether agents use resources appropriately or over-compute for limited benefit.

%These challenges are particularly acute in commercial search domains such as hotel reservation, where queries range from straightforward factual lookup to complex, multi-hop reasoning with underspecified requests. 
These challenges are especially pronounced in commercial search domains like hotel booking, where queries range from simple lookups to complex, multi-hop requests with vague constraints.
Consider two queries that illustrate this range:
(1) “\textit{Hotel with a gym in Berlin.}” A competent system can resolve location via filtering and match the amenity from structured attributes, without requiring multi-step reasoning.
(2) “\textit{A quiet, stroller-friendly boutique near Barcelona’s center with spacious rooms and step-free access, preferably one that feels authentic and not too touristy.}” %Here, the system must integrate information from both structured and unstructured sources: unstructured descriptions (quiet, boutique), structured fields (room size, accessibility tags), and underspecified constraints (“stroller-friendly” implies ramps, wide corridors, or elevators). 
The system must combine information from both structured and unstructured sources: unstructured descriptions (e.g., “quiet,” “boutique”), structured fields (room size, accessibility tags), and vague constraints like “stroller-friendly,” which could imply ramps, wide corridors, or elevators.

In this paper, we introduce \textbf{\x}~(\textbf{Hotel} \textbf{Qu}ality \& \textbf{E}fficiency \textbf{S}earch \textbf{T}estbed), a benchmark of 214 handcrafted hotel search queries, ranging from simple to complex, many of which express inherently underspecified preferences. To enable consistent and more accurate evaluation of underspecified queries, we collect clarifications -- explicit statements from query authors revealing their true intent, accessible only to the judges.
We jointly evaluate quality (relevance and factuality) and efficiency (cost and latency), analyze how query characteristics influence the behavior of lightweight retrievers and LLM-based agents, and establish an upper bound on achievable efficiency.

% \noindent\textbf{Our main contributions are:}

% \begin{enumerate}
%     \item \textbf{A benchmark for agentic search:} 214 hotel queries spanning simple to complex, with explicit complexity ratings, 
%    ground-truth clarifications for underspecified preferences, and structured decompositions enabling fine-grained analysis of agent behavior.
%     \item \textbf{Joint evaluation of effectiveness and efficiency:} Systematic evaluation that measures answer quality alongside cost, token usage, and latency, capturing the tradeoffs between quality and practical deployability.
%     \item \textbf{Empirical analysis exposing inefficiencies:} We demonstrate that, current LLM-based agents display poor cost-performance trade-offs, frequently over-investing computation for marginal quality gains. Our analysis further suggests significant potential for more cost-aware agent design.
% \end{enumerate}

\noindent\textbf{Our main contributions are:}
\\

\noindent\textbf{1. A benchmark for agentic search:} \quad
% 214 hotel queries spanning simple to complex, with explicit complexity ratings,
% ground-truth clarifications for underspecified preferences, and structured
% decompositions enabling fine-grained analysis of agent behavior.
A set of 214 simple to complex hotel queries, each with complexity ratings, ground-truth clarifications for underspecified preferences, and structured decompositions for detailed analysis of agent behavior.

\vspace{4pt}
\noindent\textbf{2. Joint evaluation of quality and efficiency:} \quad
% Systematic evaluation that measures answer quality alongside cost, token usage,
% and latency, capturing the tradeoffs between quality and practical deployability.
A systematic measurement of answer quality together with cost, token usage, and latency, capturing tradeoffs between quality and practical deployability.

\vspace{4pt}
\noindent\textbf{3. Empirical analysis exposing inefficiencies:} \quad
We demonstrate that current LLM-based agents display poor cost–quality
trade-offs, frequently over-investing computation for marginal quality gains.
Our analysis suggests significant potential for more cost-aware agent
design.

\section{Related Work}
\subsection{Benchmarks for Agentic Search}

% \sofia{I would start benchmarks for agentic search from papers that were already agentic. you could maybe say they evolved from QA and cite a few papers but use a single line - don't start with a whole paragraph about QA... so maybe just start from the second paragraph?} Early question-answering (QA) datasets like Natural Questions \cite{kwiatkowski2019natural} and TriviaQA \cite{joshi2017triviaqa}, primarily focused on simple factual queries that could be resolved through a single retrieval step or were directly encoded in a model’s parametric knowledge. As language models advanced, multi-hop QA datasets such as HotpotQA \cite{yang2018hotpotqa} and 2WikiMultiHopQA \cite{ho2020constructing} emerged, increasing task complexity by requiring models to integrate information from multiple evidence sources. However, despite this progress, these datasets still follow relatively structured and fail to capture the inherent ambiguity and exploration characteristic of real-world search.

Recent benchmarks for agentic search push beyond classical QA \cite{kwiatkowski2019natural, ho2020constructing} to multi-hop RAG \cite{tangmultihop, yang2024crag, krishna2025fact}, and further toward multi-hop reasoning and agentic research.

In the upper section of Table~\ref{tab:comparison_benchmark}, we summarize agent benchmarks spanning general \cite{gou2025mind2web, wei2025browsecomp, mialon2023gaia, andrews2025scaling}, e-commerce \cite{yao2022webshop}, and enterprise domains \cite{xu2024theagentcompany}. These works typically evaluate agents across a diverse range of tasks, involving search among other requirements, to assess their overall capabilities. The middle section of the table summarizes recent work on agentic search, highlighting that most efforts emphasize deep research \cite{du2025deepresearch, abaskohi2025drbench, rosset2025researchy}, as well as factual seeking \cite{xi2025infodeepseek} and broad search \cite{wong2025widesearch}. However, no existing work jointly evaluates efficiency and quality, nor addresses underspecified queries where implicit user intent must be inferred -- a common characteristic of real-world search that is critical for practical deployment \cite{kapoor2024ai}.

\begin{table*}[t]
\centering
\small
\caption{Comparison between benchmarks. 
Top: agentic benchmarks involving search among other requirements across general, e-commerce, and enterprise domains. 
Middle: agentic search benchmarks focusing on deep research, factual seeking, and broad search. 
Columns \textbf{A}, \textbf{F}, and \textbf{E} indicate \textbf{Accuracy}, \textbf{Factuality}, and \textbf{Efficiency}, respectively.}

\label{tab:comparison_benchmark}
\setlength{\tabcolsep}{3pt}
\begin{tabularx}{\linewidth}{@{}X p{2.1cm} r l p{2.2cm} c c c@{}}
\toprule
\textbf{Name} & \textbf{Domain} & \textbf{Size} & \textbf{Language} & \textbf{Complexity} & \textbf{A} & \textbf{F} & \textbf{E} \\
\midrule
Mind2Web 2 \cite{gou2025mind2web}
& General
& 130
& English
& High 
& $\checkmark$& $\checkmark$ & $\times$ \\

WebShop \cite{yao2022webshop}
& E-Commerce 
& 12,087
& English 
& Low 
& $\checkmark$ & $\times$ & $\times$ \\

BrowseComp \cite{wei2025browsecomp}
& General 
& 1,266
& English 
& High 
& $\checkmark$ & $\times$ & $\times$ \\

TheAgentCompany \cite{xu2024theagentcompany}
& Enterprise 
& 175
& English 
& Undefined 
& $\checkmark$ & $\times$ & $\checkmark$ \\

GAIA \cite{mialon2023gaia}
& General assistant  
& 466
& English 
& Low to High
& $\checkmark$ & $\times$ & $\times$ \\

GAIA2 \cite{andrews2025scaling}
& General 
& 963
& English 
& High
& $\checkmark$ & $\checkmark$ & $\times$ \\

%\addlinespace
\midrule

InfoDeepSeek \cite{xi2025infodeepseek}
& Search
& 245 
& 19 languages
& High
& $\checkmark$ & $\checkmark$ & $\times$ \\

DeepResearch Bench \cite{du2025deepresearch}
& Research
& 100
& English ; Chinese
& High 
& $\checkmark$ &$\checkmark$ & $\times$ \\

LiveDRBench \cite{java2025characterizing}
& Research
& 100
& English 
& High 
& $\checkmark$ & $\times$ & $\times$ \\

WideSearch \cite{wong2025widesearch} 
& Search
& 200
& English ; Chinese 
& Medium 
& $\checkmark$ & $\checkmark$ & $\times$ \\

%\addlinespace
\midrule
\textbf{ \x} (Ours)
& Hotels
& 214
& English
& Low to High
& $\checkmark$ & $\checkmark$ & $\checkmark$ \\
\bottomrule
\end{tabularx}
\end{table*}

\subsection{Efficiency in LLMs and Agents}

% The growing use of test-time compute to enhance reasoning and performance \cite{jaech2024openai, snell2024scaling}, exemplified by methods such as Chain-of-Thought prompting \cite{wei2022chain}, has underscored the importance of computational efficiency. While advances in reasoning capabilities of LLMs and agentic systems have led to impressive gains across a wide range of tasks \cite{ferrag2025llm}, they have also introduced substantial computational costs, often in ways that are unnecessary or impractical for real-world applications \cite{feng2025efficient}. 

Recent work explores ``fast'' and ``slow'' thinking in LLMs \cite{kahneman2011thinking, wang2025harnessing}. Slow thinking uses test-time compute to enhance reasoning \cite{jaech2024openai, snell2024scaling}, exemplified by Chain-of-Thought \cite{wei2022chain}. %While these methods achieve impressive gains \cite{ferrag2025llm}, they introduce computational costs that are often impractical for real-world applications \cite{feng2025efficient}. 
Although these methods deliver strong gains~\cite{ferrag2025llm}, they often incur computational costs that are impractical for real-world use~\cite{feng2025efficient}.
%However, 
Moreover, current LLMs lack the ability to adaptively choose between these modes. Using fast thinking on complex queries degrades quality, while applying slow thinking to simple queries wastes computational resources. 

Recent work proposes hybrid frameworks for adaptive mode selection \cite{jiang2025think, fang2025thinkless, cheng2025incentivizing}, yet existing benchmarks remain limited, not specifically designed for agentic search or efficiency–quality trade-offs. With the rise of search agents \cite{zhang2025deep}, the problem has become more pronounced, as their extended reasoning traces often lead to computationally intensive processes for completing complex tasks \cite{xu2025comprehensive, li2025webthinker}.

To the best of our knowledge, no existing benchmark systematically evaluates this capability in agentic search. Therefore, we propose a new benchmark designed to fill this gap and enable rigorous evaluation in commercial contexts.

% Although research on models capable of integrating both “slow” and “fast” thinking has advanced in recent years \cite{kahneman2011thinking, jiang2025think, wang2025harnessing}, there remains a substantial gap in making such models practical for real-world commercial settings. To the best of our knowledge, no existing benchmark systematically evaluates this capability. Therefore, we propose a new benchmark designed to fill this gap and enable rigorous evaluation in commercial contexts.

% \sofia{we might need to add some references for vagueness and query implicitness awareness} \guy{for 2.1 or 2.2? }\sofia{2.1 probably. or its own sub-section but only if you super shorten everything else.}

\section{The \xspace Benchmark}
\subsection{Problem Definition}
% \haggai{these headers are a waste of space and doesn't really contribute to clarity}
%\paragraph{Setting.}
Let $\mathcal{H}=\{h_1,\dots,h_N\}$ denote a hotel catalog.
Given a natural-language query $q\in\mathcal{Q}$, we extract a finite set of \emph{qualifiers} (constraints)
$\Phi(q)=\{\varphi_1,\ldots,\varphi_m\}$ over attributes such as location, budget, amenities, etc.
The task is to retrieve the top-$k$ relevant hotels to $q$. For generative models, the output should include grounded evidence, which justifies the reasoning behind its selections.
% The task is to retrieve $L(q)$ -- the top-$k$ relevant hotels to $q$. When the model produces a generative response, it also provides supporting evidence \(A(q)\), which explains or justifies the reasoning behind its selections. \ran{the last sentence is not clear} \guy{better?}
\subsection{Query Collection}

Twenty-two human annotators participated in the data creation process, guided by a three-stage protocol designed to ensure diversity in complexity and query characteristics. An additional human reviewer then filtered out queries that did not adhere to the task guidelines, ensuring that only well-formed and goal-oriented queries were retained. % in the benchmark.

\paragraph{Stage 1: Query generation.} 

Annotators wrote queries based on authentic travel scenarios they would realistically search for. We instructed them to express their requirements as they naturally would when using a natural language search interface. This yielded queries spanning simple lookups to complex and multi-constraint requests, reflecting real-world patterns where users leverage natural language interfaces rather than traditional keyword or filter-based interfaces.

\paragraph{Stage 2: Clarification ground truth.}
\label{sec:clarification}
Each annotator also provided a \textbf{clarification}--a note that makes their underspecified assumptions explicit. This gap is evident in our query analysis and aligns with prior observations in the literature~\cite{choi2025bloomintent, dou2007large}.

This step is motivated by a central insight from \citet{thomas2024large}:
\emph{the only reliable “gold” relevance signal is the intent of the searcher themselves}. The goal is to capture what a capable agent must infer to correctly interpret the request. Clarifications are only available to the \emph{judge}, and they serve as ground truth for the user's implicit intent. 

Clarifications can take many forms. For example, an underspecified request like “\textit{Hotel for a solo traveler}” is clarified as “\textit{Find affordable hotels or hostels in safe neighborhoods suitable for solo travelers.}” Similarly, \textit{“Hotels in London where I can see the King”} can be clarified by specifying the location being referenced,  for instance, indicating that it refers to Buckingham Palace in London.

\paragraph{Stage 3: Complexity assessment.}

Annotators rated the \textbf{complexity} of each query as \emph{Simple}, \emph{Moderate}, or \emph{Complex}. The annotators' complexity assessments are guided by the following three-level rubric:
\begin{itemize}
    \item \textbf{1 = Simple:} solvable within approximately 5 minutes of search.
    \item \textbf{2 = Moderate:} requires roughly 5 to 15 minutes of exploration.
    \item \textbf{3 = Complex:} involves multi-step reasoning, cross-referencing, or multi-source search, typically exceeding 15 minutes.
\end{itemize}
This time-based interpretation of query complexity follows prior work showing that human solution time correlates with task difficulty~\cite{gou2025mind2web}, and relies on the established assumption that users can reliably self-assess the informational needs of their queries~\cite{suri2024use}.

% See Appendix~\ref{apndx:benchmark} for further details.

\subsection{Query Characterization}
\label{sec:dataset-characterization}

% \paragraph{Overview.} 
Our dataset consists of \textbf{214 queries}, out of which \textbf{73.4\%} include a clarification. 
The complexity distribution shows 37.8\% are labeled Complex, 37.4\% Moderate, and 24.8\% Simple, providing balanced coverage across difficulty levels.

% \paragraph{Query decomposition.}
To enable fine-grained analysis of our benchmark, we decompose each query $q$ into a set of subqueries $\{q_i\}$, where each $q_i$ corresponds to a distinct \textit{qualifier} capturing a specific aspect of user intent. For example:

\begin{quote}
\textit{``I'm going for a solo trip to San Jose, Costa Rica. Find me a hotel with great social atmosphere.''}
\end{quote}

This query contains three pairs of qualifiers: \textit{``Solo trip''} (explicit, \textit{Population}), \textit{``San Jose, Costa Rica''} (explicit, \textit{Location}), and \textit{``Great social atmosphere''} (implicit, \textit{Description}).
% This query contains three qualifiers:
% \begin{itemize}
%     \item \textbf{``Solo trip''}: Explicit \textit{Population} qualifier
%     \item \textbf{``San Jose, Costa Rica''}: Explicit \textit{Location} qualifier  
%     \item \textbf{``Great social atmosphere''}: Implicit \textit{Description} qualifier
% \end{itemize}
We annotate each qualifier along two dimensions: \textbf{Type} (e.g., Explicit vs. Implicit, Negation) and \textbf{Content} (e.g., Location, Population, Description). This taxonomy was iteratively derived  by multiple annotators analyzing an initial subset of queries (see Table~\ref{tab:qualifiers} for the complete taxonomy with examples).

% This decomposition allows us to analyze how specific query characteristics, such as the number of qualifiers, their explicitness, or content type, affect model quality and efficiency across different system architectures.
This decomposition lets us examine how query features such as the number of qualifiers, their explicitness, and content type influence model quality and efficiency across system architectures.

\subsection{Hotels Corpus}
We use two complementary data sources: the first is a large collection of textual \textit{hotel descriptions} covering approximately one million hotels\footnote{\url{https://www.kaggle.com/datasets/raj713335/tbo-hotels-dataset}} and the second is \textit{HotelRec} \cite{antognini2020hotelrec}, a large-scale hotel recommendation dataset derived from TripAdvisor containing around 50 million user reviews. We retain only reviews corresponding to hotels for which a textual description is available. After preprocessing, we obtain \textbf{963,028} hotel descriptions. The adapted review dataset comprises \textbf{21,112,546} reviews covering \textbf{106,239} unique hotels, \textbf{18,520} cities, and \textbf{132} countries. Each hotel has \textbf{1} to \textbf{31,219} reviews, with a median of \textbf{68.0} and a mean of \textbf{198.7}. The full description of the indexing setup is presented in Appendix~\ref{app:indexing}.

% \section{Experiments and Analyses}
\section{Experimental Setup}

\begin{table*}[t]
\centering
\small
\setlength{\tabcolsep}{3pt}
\begin{tabular}{@{}p{2.2cm}p{2.8cm}lcc|cccc@{}}
\toprule
\textbf{Section} & \textbf{Model} & \textbf{Subset} & \multicolumn{2}{c|}{\textbf{Quality}} & \multicolumn{4}{c@{}}{\textbf{Efficiency}} \\
\cmidrule(lr){4-5} \cmidrule(lr){6-9}
& & & \textbf{Accuracy} & \textbf{Factuality} & \textbf{Cost (\$)} & \textbf{\#Tokens} & \textbf{P50 (s)} & \textbf{P90 (s)} \\
\midrule

\multirow{6}{2.2cm}{\textbf{Retrieval only}}
& \multirow{2}{2.8cm}{BM25} & \textit{Reviews} & 2.64 & -- & 0.00 & -- & 0.23 & 0.23 \\
& & \textit{Descriptions} & 1.80 & -- & 0.00 & -- & 0.0046 & 0.0046 \\
\cmidrule{2-9}
& \multirow{2}{2.8cm}{Dense (22M)} & \textit{Reviews} & 2.56 & -- & 0.00 & -- & 0.0007 & 0.0007 \\
& & \textit{Descriptions} & 2.22 & -- & 0.00 & -- & 0.0087 & 0.0087 \\
\cmidrule{2-9}
& \multirow{2}{2.8cm}{Dense (300M)} & \textit{Reviews} & 3.00 & -- & 0.00 & -- & 0.0054 & 0.0054 \\
& & \textit{Descriptions} & 2.63 & -- & 0.00 & -- & 0.0169 & 0.0169 \\
\midrule

\multirow{4}{2.2cm}{\textbf{Retrieval + LLM Reranker}}
& \multirow{2}{2.8cm}{Dense (300M) + Reranker (600M)} & \textit{Reviews} & 3.26 & -- & 0.61 & -- & 2.9511 & 3.7701 \\
& & \textit{Descriptions} & 2.77 & -- & 0.76 & -- & 3.6254 & 4.5331 \\
\cmidrule{2-9}
& \multirow{2}{2.8cm}{Dense (300M) + Reranker (4B)} & \textit{Reviews} & 3.32 & -- & 3.31 & -- & 16.070 & 19.7119 \\
& & \textit{Descriptions} & 2.96 & -- & 4.02 & -- & 19.2011 & 24.4993 \\
\midrule

\multirow{4}{2.2cm}{\textbf{LLM-based Agents}}
& Qwen3-32B & \textit{Full} & 3.82 & 2.43 & 4.45 & 13M/3M & 115.74 & 161.93 \\
& Claude 4.5 Haiku & \textit{Full} & 3.57 & 2.81 & 18.92 & 1M/0.2M & 69.40 & 155.32 \\
& Claude 3.7 Sonnet & \textit{Full} & 4.22 & 2.97 & 96.03 & 14M/3.5M & 364 & 938.42 \\
& Claude 4 Sonnet & \textit{Full} & 4.11 & 2.83 & 50.16 & 7.9M/1.8M & 123.44 & 291.76 \\

\midrule
    \multicolumn{2}{@{}l}{\textbf{Budget Oracle \$1}} & \textit{Full} & 4.23 & -- & 1.00 & -- & 22.58 & 31.55 \\

    \multicolumn{2}{@{}l}{\textbf{Budget Oracle \$2}} & \textit{Full} & 4.42 & -- & 1.94 & -- & 32.13 & 44.68 \\

        \multicolumn{2}{@{}l}{\textbf{Budget Oracle \$4}} & \textit{Full} & 4.55 & -- & 3.99 & -- & 37.70 & 57.14 \\
\multicolumn{2}{@{}l}{\textbf{Quality Oracle}} & \textit{Full} & 4.71 & -- & 13.10 & -- & 62.65 & 127.44 \\

% \addlinespace[2pt]
% \multicolumn{9}{@{}l}{\emph{Oracle selects, per query, the cheapest model among those tied for best quality.}}

% \bottomrule
\end{tabular}
\caption{Evaluation split into \textbf{Retrieval only}, \textbf{Retrieval + LLM-based Reranker}, and \textbf{LLM-based Agents} on \textbf{Reviews} and \textbf{Descriptions}, as well as two versions of \textbf{Oracle} models. Metrics cover \textbf{Quality} and \textbf{Efficiency}.}
\label{tab:results}
\end{table*}

\paragraph{Models.}\label{sec:models}
We evaluate baselines spanning the quality-efficiency spectrum: from fast, lightweight retrieval methods to sophisticated but costly LLM-based agents, for the task of returning the top-3 hotels for each query. For retrieval baselines, we employ BM25 \cite{lu2024bm25s} and top-performing embedding models from the \textit{MTEB} benchmark \cite{muennighoff2023mteb}\footnote{\url{https://huggingface.co/spaces/mteb/leaderboard}} in two size categories: \texttt{all-MiniLM-L6-v2} (22M parameters) \cite{wang2020minilm} and \texttt{embeddinggemma-300m} (300M parameters) \cite{vera2025embeddinggemma}.

As additional baselines with a reranking stage, we incorporate an LLM reranker that estimates the probability of answering ``Yes'' to the question of whether a given document is relevant to the query. Specifically, we employ \texttt{Qwen3-Reranker-0.6B} and \texttt{Qwen3-Reranker-4B} \cite{zhang2025qwen3}. Each retriever is evaluated separately on both databases, reviews and descriptions. %For additional details regarding the retrieval baselines, see Appendix~\ref{sec:ret}.
For more details on the retrieval baselines, see Appendix~\ref{sec:ret}.

For agentic baselines, we utilize Claude models (Sonnet 4, Sonnet 3.7, and Haiku 4.5) \cite{anthropic2025claude37, anthropic2025claude4, anthropic2025claude45} and Qwen3-32B \cite{yang2025qwen3} within the LangGraph framework\footnote{\url{https://www.langchain.com/langgraph}}. Each agent orchestrates three information sources: hotel \textit{Descriptions}, customer \textit{Reviews}, and \textit{Web Search} via the Tavily API\footnote{\url{https://www.tavily.com}}, following the iterative workflow described below.

\paragraph{Agentic workflow.} The agent operates through an iterative process (Figure~\ref{fig:agent}) for \(t = 1, \dots, T\) with memory state \(m_t\) (a textual summary of hotels retrieved so far) consisting of:  
\emph{(i) Plan:} select a source \(s_t \in S = \{\text{Descriptions}, \text{Reviews}, \text{Web Search}\}\) and generate a search query \(r_t\) based on the original query \(q\) and memory \(m_{t-1}\);  
\emph{(ii) Retrieve:} execute query \(r_t\) on source \(s_t\) to fetch up to \(k\) hotel candidates \(H_t \subseteq \mathcal{H}\);  
\emph{(iii) Filter:} prune irrelevant results from \(H_t\) and update memory to \(m_t\) with newly found hotels.  
The loop terminates when \(k\) hotels are identified or \(T\) has been reached, yielding the final ranked list with grounded evidence. For more details about the agent, see Appendix \ref{sec:agnt}.

\paragraph{Oracle models.} Finally, to quantify the potential for improvement, we introduce two oracle baselines representing upper bounds on achievable quality. The \textbf{budget oracle} maximizes overall accuracy under fixed budget constraints (e.g., \$1, \$2, and \$4), formulated as a Multiple-Choice Knapsack problem \cite{sinha1979multiple}. The \textbf{quality oracle} selects, per query, the cheapest model achieving the highest accuracy.

\paragraph{Evaluation.}
We evaluate the baselines along two complementary axes: \textit{quality} and \textit{efficiency}. For quality, we employ an LLM-as-a-judge approach to assess: (i)~\textit{accuracy}, which measures how well the answer aligns with the user's requirements, and (ii)~\textit{factuality}, which measures how well it is grounded in retrieved data with proper citations. Both metrics use a scoring guideline with well-defined criteria for assigning scores from 1 to 5, as shown in Appendix Table~\ref{tab:hotel-rubric} (details in Appendix~\ref{apndx:eval} and \ref{apndx:factuality}). We use Sonnet~4.5 \cite{anthropic2025sonnet45} as the judge model. To ensure consistent evaluations and address query underspecification, we provide the LLM judge with the \textit{Clarification} from Section~\ref{sec:clarification}, which captures the annotator's true intent. We validate this approach by measuring agreement between LLM and human evaluators on 246 answers spanning all baseline types, achieving a weighted Cohen's kappa of 0.84. For more details about agreement evaluation, see Appendix \ref{app:agreement}.

For efficiency, we measure the total number of tokens processed (input/output), the cost of API usage\footnote{All costs are based on Amazon Bedrock pricing as of November~2025.}, and latency statistics, specifically the median (\textbf{P50}) and tail (\textbf{P90}) response times. These metrics jointly capture the trade-off between model capability and practical deployability in real-world scenarios. For more details, see Appendix \ref{app:agreement}.

\begin{figure*}[t]
\centering
\includegraphics[width=\linewidth]{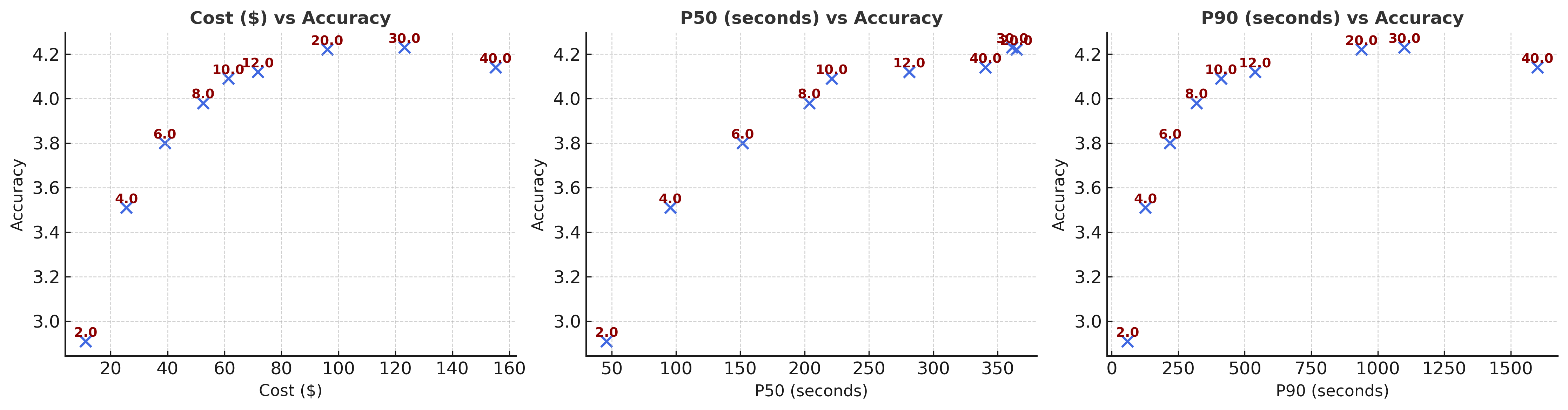}
\caption{\textbf{Accuracy--Efficiency Trade-off.} Numbers above each point indicate the agent’s iteration limit. \textbf{Left:} As cost increases, accuracy initially improves, but beyond a certain point additional cost yields no further gains. \textbf{Middle:} A similar pattern appears with median latency: accuracy rises with longer deliberation until both metrics plateau and converge. \textbf{Right:} The P90 latency curve mirrors the cost trend, indicating that on some queries the model fails to terminate early, leading to disproportionately high latency and cost.}
\label{fig:cost-inefficiency}
\end{figure*}

\section{Results and Analysis}
\subsection{Quality \& Efficiency Comparison}

Table~\ref{tab:results} presents the main evaluation results on \x, comparing retrieval-based baselines with LLM-based agentic systems. Retrieval methods (BM25, dense retrievers)  offer near-zero cost and latency but have limited reasoning capabilities, resulting in lower overall accuracy compared to highly capable LLM-based agents. Among all models, Sonnet 3.7 achieves the highest accuracy but is also significantly more expensive (see Section~\ref{subsection:Cost Inefficiency Analysis} for detailed analysis and discussion). 

The results reveal a substantial quality-efficiency gap: retrieval models excel in cost efficiency, while advanced LLMs lead in accuracy. This gap constrains deployment in industrial search pipelines where latency, scalability, and cost are critical, underscoring the need for efficient agentic architectures that deliver strong quality without high computational overhead.

\begin{figure}[t]
\centering
\includegraphics[width=1\linewidth]{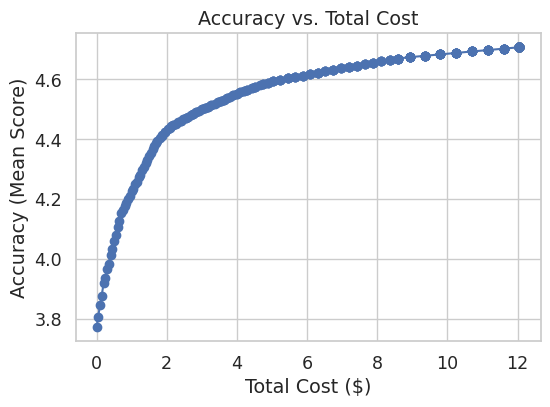}
\caption{Budget Oracle. Accuracy achieved by solving a multiple-choice knapsack problem under varying budget limits. A clear elbow appears around \$2.}
\label{fig:oracle_1}
\end{figure}

\paragraph{Oracle Baselines.} To establish theoretical upper bounds on routing efficiency, we evaluate two oracle strategies with perfect foresight. The budget oracle formulates model selection as a multiple-choice knapsack problem: given a global monetary budget, it selects exactly one model per query to maximize total accuracy without exceeding the budget constraint. As shown in Figure~\ref{fig:oracle_1}, accuracy exhibits a clear elbow around \$2, beyond which additional expenditure yields diminishing returns.

The quality oracle operates per-query, selecting the cheapest model among those achieving the highest accuracy, thereby providing an upper bound on ideal routing given perfect knowledge of query difficulty. Figure~\ref{fig:oracle_2} reveals that most queries achieve optimal accuracy using relatively inexpensive models, with only a small fraction requiring the most powerful agents. Both oracles demonstrate that near-optimal accuracy is attainable at a fraction of current costs: the quality oracle outperforms the best agent at lower cost, while the budget oracle at \$1 achieves higher accuracy than all agents while costing 96× less than Sonnet 3.7 and 4× less than Qwen3-32B. 

These results reveal substantial headroom for adaptive routing strategies and suggest that heavy agentic reasoning is rarely necessary, with large models delivering outsized benefits only on a minority of challenging queries.

\subsection{Cost Inefficiency Analysis}
\label{subsection:Cost Inefficiency Analysis}

Tracing agent reasoning trajectories reveals notable inefficiencies. Agents often continue invoking tool calls after retrieving sufficient evidence, repeatedly issuing nearly identical searches despite having access to their search history.

To quantify this, we limit the number of tool invocations for Sonnet 3.7 and measure the impact on quality and efficiency. Figure~\ref{fig:cost-inefficiency} shows that excessive tool calls lead to over-exploration: increased cost and latency without corresponding gains in accuracy, as median latency remains stable. 

These findings highlight a critical gap: the lack of cost-aware stopping criteria in current agentic architectures. Potential solutions include contract algorithms \cite{shmueli2009best}, learned stopping policies \cite{yuan2024following}, and RL-based resource allocation \cite{aggarwal2025l1}.

\subsection{Influence of Query Qualifiers} \label{sec:query-features}

We examine how query attributes influence model quality using the taxonomy from Section~\ref{sec:dataset-characterization}, extended with query length and human-rated complexity. We apply Welch's \textit{t}-test or Spearman correlation depending on feature type, retaining only significant results ($\alpha < 0.05$). Table~\ref{fig:features} in Appendix~\ref{apndx:queries_analysis} presents the complete analysis across all models and query attributes.

\paragraph{Retrieval vs. agentic models.} Query complexity significantly affects retrieval-based models but not agentic models, leading to less accurate answers that fail to fully satisfy user requirements. Query length influences both retrieval models and smaller agents like Qwen3-32B. Qwen3-32B is also sensitive to the number of qualifiers and linguistic properties like negation and subjectivity, which further decrease response accuracy. See Appendix~\ref{apndx:queries_analysis} for the full results. 

% \paragraph{Agent behavior across complexity levels.} We further analyze how agents respond to query complexity. Qwen3 and Sonnet 4 exhibit a monotonic increase in cost across all complexity levels, Haiku only between simple and non-simple queries, while Sonnet 3.7 shows the opposite trend. In terms of latency, Qwen3, Sonnet 4, and Haiku are monotonic across all levels, whereas Sonnet 3.7 again diverges. For accuracy, Qwen3, Sonnet 4, and Sonnet 3.7 maintain monotonic trends between simple and non-simple queries, with Haiku being the only model that does not. These findings suggest that while larger and more capable agents scale predictably with query complexity, smaller or less stable models exhibit inconsistent behavior, highlighting opportunities for adaptive cost–quality routing. see Appendix~\ref{sec:complx} for the full results. 

\paragraph{Agent behavior across complexity levels.}
We further analyze how agents respond to query complexity. Qwen3-32B and Sonnet~4 increase cost, latency, and token usage as complexity rises, indicating that they invest more computation in harder queries. Haiku also spends more, but mainly when moving from simple to non-simple queries, with a slight cost drop at the highest level. In contrast, Sonnet~3.7 uses \emph{less} cost, latency, and tokens as complexity increases, suggesting miscalibrated stopping behavior. Accuracy is highest on simple queries for all models and generally drops on more complex ones, with only partial recovery at the highest level. Overall, most agents respond to complexity by doing more work, but this extra effort only partially offsets the accuracy degradation on harder queries, while Sonnet~3.7 appears under-invested exactly where queries are most difficult. See Appendix~\ref{apndx:analysis} for full results.

This analysis primarily reflects the pre-retrieval stage, capturing how query properties (e.g., length, specificity, etc.) a priori affect the system’s ability to retrieve relevant evidence, rather than its subsequent reasoning or answer-generation processes \cite{roitman2020ictir}.

% % ----- another proposal to replace the above paragraph ---- %
% We further analyze how agents respond to query complexity. Models exhibit distinct scaling patterns across cost, latency, and accuracy (Figure~\ref{tab:complexity}):

% \textbf{Cost:} Qwen3 and Sonnet 4 show monotonic increases in cost and latency across all complexity levels (Simple → Moderate → Complex), while Haiku increases only between simple and non-simple queries. Sonnet 3.7 exhibits the opposite trend, with costs decreasing as complexity increases.

% \textbf{Latency:} Qwen3, Sonnet 4, and Haiku scale monotonically with complexity, whereas Sonnet 3.7 again shows inverse behavior.

% \textbf{Accuracy:} Qwen3, Sonnet 4, and Sonnet 3.7 decrease monotonically from simple to non-simple queries, while Haiku shows no consistent pattern.

% Retrieval methods such as BM25 and dense retrievers achieve near-zero cost and latency but offer limited reasoning capability,  leading to lower overall performance  compared to highly-capable LLM-based agents. %\haggai{no need to point reader to this}%Section~\ref{sec:query-features} analyzes how query characteristics influence these differences.

% The results reveal a substantial gap along the quality-efficiency spectrum: retrieval models excel in cost-effectiveness, while advanced LLMs dominate in accuracy and factuality. This gap limits deployment in industrial search flows, where latency, scalability, and cost are critical constraints, underscoring the need for more efficient agentic architectures that achieve strong performance without high computational overhead.

\section{Conclusion}
\label{sec:conclusion}

We have introduced \x, a benchmark designed for evaluating hotel search agents through a diverse set of manually-written queries ranging from simple to complex, often containing inherently underspecified dimensions. To mitigate ambiguity in user intent, we incorporated explicit clarifications within our evaluation framework, ensuring more reliable and interpretable evaluations.
Our experiments span lightweight and cost-efficient retrieval models up to large LLM-based agents that demonstrate higher reasoning capabilities at the expense of latency and cost. We have further analyzed factors that influence model behavior in this setting, including the agent’s stopping decisions and the impact of linguistic and semantic features of queries on model performance. Overall, our study highlights a critical gap between quality and efficiency, underscoring the need for future research on joint optimization strategies that balance response quality with computational and economic cost.

\section{Limitations}
To ensure realism and reduce annotation bias, annotators were not exposed to any specific hotels or label sets when composing queries. This design encourages natural, diverse, and unconstrained formulations. However, it also introduces uncertainty: we cannot guarantee that a single objectively optimal answer exists for every query, nor can we precisely characterize the upper bound of achievable quality.

% Due to the inherent nondeterminism of large proprietary LLMs \cite{atil2024non}, exact reproducibility of results cannot be guaranteed. Differences in agentic workflows, execution traces, and generation trajectories may lead to variance in both output quality and computational efficiency across runs.
Because large proprietary LLMs are inherently nondeterministic~\cite{atil2024non}, exact reproducibility is not guaranteed. Variations in agent workflows, execution traces, and generation trajectories can lead to differences in both output quality and computational efficiency across runs.

% As in other LLM-prompting studies \cite{chen2024style}, our results may also be sensitive to prompt phrasing and structure. Although our prompts underwent extensive review and iterative refinement, prompt optimization for this task remains an open challenge and a promising direction for future research.
As in other LLM-prompting studies~\cite{chen2024style}, our results may be sensitive to prompt wording and structure. Although we extensively reviewed and refined our prompts, optimizing them for this task remains an open challenge and a promising direction for future work.

Finally, similar to other human-authored query benchmarks in the field, our dataset contains a relatively limited number of queries. While this reflects the substantial cost for high-quality human annotation, it may constrain statistical power and should be considered when interpreting aggregate quality metrics.

\section{Ethics Statement}

During our data filtering process, we proactively removed all queries containing offensive, inappropriate, or harmful language to ensure the safety and integrity of the dataset. Based on these procedures, we believe that the resulting benchmark poses minimal risk and is unlikely to produce negative societal impacts.
All language models used in this work were accessed via the Hugging Face Hub \cite{wolf2020transformers} and Amazon Bedrock. We only utilized models whose licenses explicitly permit research use, and we adhered to all relevant terms of service and usage policies throughout our experiments.
We conducted our study in accordance with standard ethical principles for data handling, model usage, and reproducibility in NLP research.

\bibliography{custom}

\clearpage

\appendix

\section{Additional Details on the HotelQuEST Benchmark}
\label{apndx:benchmark}

\subsection{Query Length}
\label{apndx:overview}

\begin{figure}[tbh]
\centering
\includegraphics[width=\linewidth]{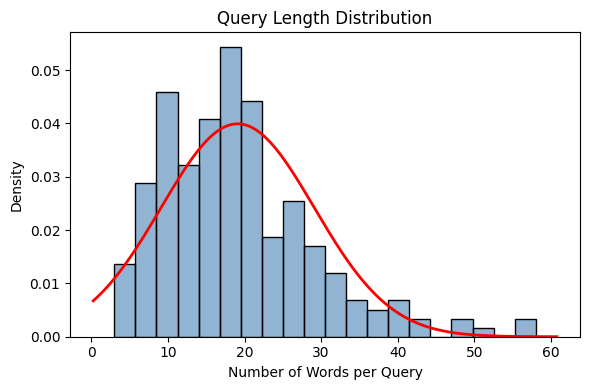}
\caption{Distribution of query lengths.}
\label{fig:dist_len}
\end{figure}

Figure~\ref{fig:dist_len} presents the distribution of query lengths, shown as a histogram over the number of words per query. As observed in our analysis, query length plays a significant role, particularly for retrieval-only models, whose performance is more sensitive to shorter and less informative queries.

\begin{table}[h]
\centering
\small
\renewcommand{\arraystretch}{1.05}
\setlength{\tabcolsep}{6pt}
\begin{tabular}{@{}ll@{}}
\toprule
\textbf{Qualifier Type} & \textbf{Qualifier Content} \\ 
\midrule
Explicit / Implicit & Purpose \\
Negation & Location \\
Similarity & Population \\
Range & Seasonality \\
Time-sensitive & Description \\
Optional / Mandatory & Rating \\
\bottomrule
\end{tabular}
\caption{Taxonomy of qualifier types and contents.}
\label{tab:qualifiers}
\end{table}

% \subsection{Complexity}
% \label{sec:complx}

% \subsection{Query Complexity}
% \label{apndx:complexity}

\begin{figure*}[h]
\centering
\includegraphics[width=1.0\linewidth]{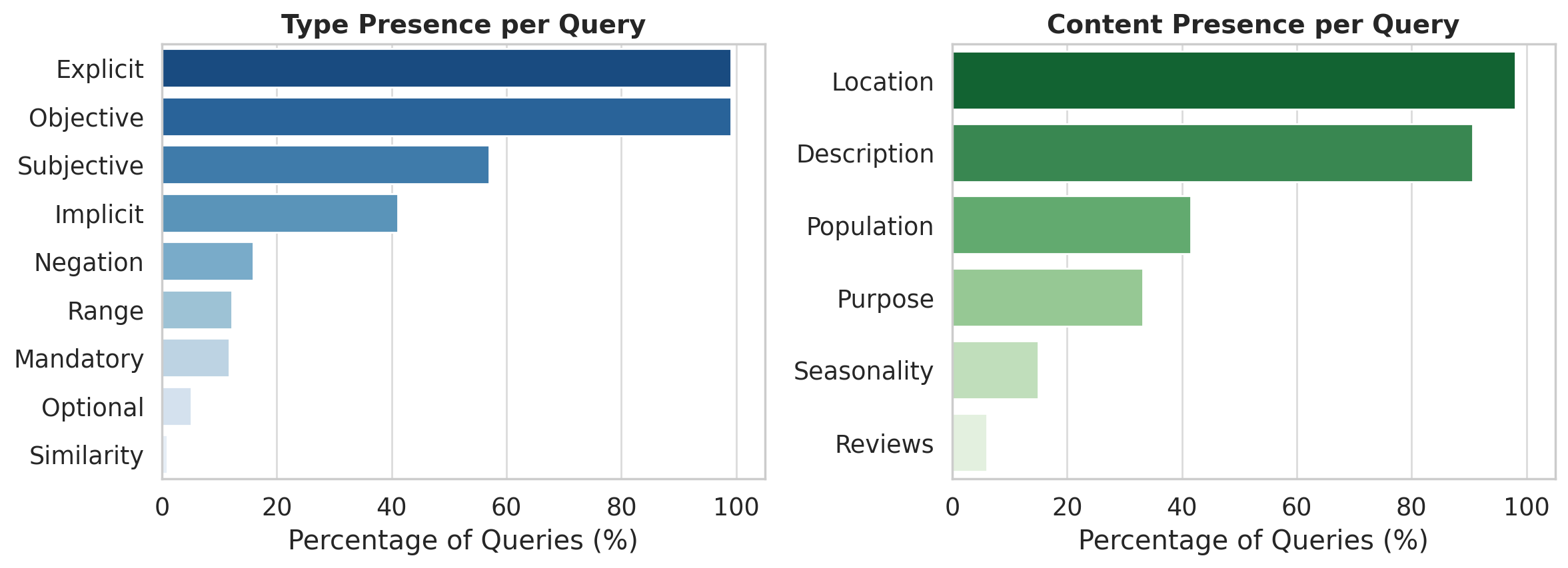}
\caption{Analysis of the queries by the presence of qualifier attributes.}
\label{fig:dist}
\end{figure*}

\section{Additional Details on the Experiments}
\label{apndx:exps}

\guy{adding the prompt for the judges}

\subsection{Judgment}
\label{sec:judge}

All judgments in this work are produced using \textit{Claude Sonnet 4.5} \cite{anthropic2025sonnet45} as the evaluating model for his strong performance \cite{zheng2023judging}. We employ two dedicated prompts: the \emph{accuracy} prompt (see \ref{apndx:acc}) and the \emph{factuality} prompt (see \ref{apndx:factuality}). These prompts provide structured scoring criteria to ensure consistent and reproducible evaluations across all model outputs.

\subsection{Agent Workflow}
\label{sec:agnt}

The agent operates with three specialized tools: one for retrieving item descriptions, one for retrieving reviews, and one for performing web search. After each tool call, the agent extracts only the information relevant to the user query and stores it in an internal notes field. This mechanism prevents repeated regeneration of long, irrelevant context across iterations and ensures that the model accumulates only the essential evidence needed for reasoning.

\begin{figure*}[h]
\centering
\includegraphics[width=0.8\linewidth]{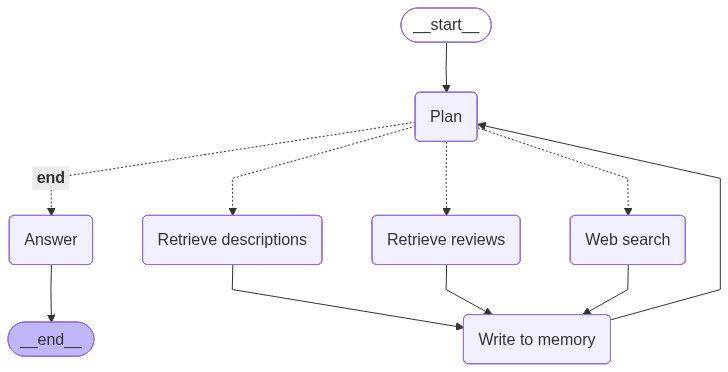}
\caption{Illustration of the agentic workflow.}\label{fig:agent}
\end{figure*}

Figure~\ref{fig:agent} illustrates the full agentic workflow. At the beginning of each episode, the agent receives the user query and decides whether to (i) call a tool or (ii) generate a final answer. When a tool is selected, the retrieved information is summarized and added to the notes, after which the agent replans its next step. This iterative process continues until the agent determines it has sufficient evidence and produces the final answer.

\begin{table*}[t]
\centering
\caption{Metrics by complexity level and model; tokens shown as inputK/outputK.}
\resizebox{\textwidth}{!}{
\begin{tabular}{l *{12}{r}}
\toprule
& \multicolumn{4}{c}{\textbf{Simple}} & \multicolumn{4}{c}{\textbf{Moderate}} & \multicolumn{4}{c}{\textbf{Complex}} \\
\cmidrule(lr){2-5}\cmidrule(lr){6-9}\cmidrule(lr){10-13}
\textbf{Metric} & Qwen3-32B & Sonnet 4 & Haiku 4.5 & Sonnet 3.7 & Qwen3-32B & Sonnet 4 & Haiku 4.5 & Sonnet 3.7 & Qwen3-32B & Sonnet 4 & Haiku 4.5 & Sonnet 3.7 \\
\midrule
\textbf{Cost}
& 0.022 & 0.212 & 0.078 & 0.482
& 0.025 & 0.226 & 0.095 & 0.479
& 0.026 & 0.250 & 0.092 & 0.461 \\
\textbf{Tokens}
& 77K/18K & 33K/8K & 31K/8K & 73K/18K
& 83K/20K & 36K/8K & 37K/10K & 73K/17K
& 90K/21K & 39K/9K & 35K/10K & 70K/17K \\
\textbf{Latency (sec)}
& 83.78 & 136.60 & 74.36 & 425.95
& 89.99 & 141.24 & 91.42 & 423.81
& 96.98 & 156.88 & 93.02 & 416.82 \\
\textbf{Accuracy}
& 4.135 & 4.275 & 3.519 & 4.423
& 3.613 & 3.974 & 3.613 & 4.150
& 3.850 & 4.093 & 3.550 & 4.175 \\
\bottomrule
\label{tab:complexity_models_metrics}
\end{tabular}}
\end{table*}

\noindent\textbf{Hardware.}
Inference latency and monetary cost are evaluated on Amazon EC2 instances. For the LLMs, we employ the Amazon Bedrock API as the serving environment. For the rerankers and retrieval components, we run all computations directly on the same EC2 machine type \texttt{g6e.4xlarge} to ensure consistent quality measurement across models.

\noindent\textbf{Indexing.} We construct separate vector indices for the descriptions and the reviews using \textit{Milvus} \cite{wang2021milvus} with \textit{All-MiniLM-L6-v2} embeddings. For hotel descriptions, we adopt a \textit{FLAT} index to enable exact similarity search, while for reviews we use an \textit{HNSW} \cite{malkov2018efficient} index to improve computational efficiency at scale. \label{app:indexing} 

\subsection{Retrieval Baselines}
\label{sec:ret}

For all retrieval baselines, we rely on publicly available models from the Hugging Face Hub and the \texttt{sentence-transformers} library. All embedding models and rerankers are used in their original form without additional fine-tuning. For \textit{EmbeddingGemma}, we also adopt the prompt templates recommended by the authors to ensure consistent embedding behavior.

To index the corpora, we use \textit{FLAT} for the hotel-description collection and \textit{HNSW} for the reviews corpus. This choice is driven by computational constraints: the reviews corpus is too large for brute-force nearest-neighbor search, making hierarchical indexing essential for tractable retrieval.  
Importantly, the difference in indexing structures also explains the observed latency differences. Despite being a significantly larger corpus, the reviews collection benefits from the efficiency of \textit{HNSW}, resulting in lower latency compared to FLAT. In contrast, for BM25 we observe the opposite trend---the smaller corpus yields faster retrieval, as expected under inverted-index search.

For reranking-based baselines, we first retrieve the top 100 documents from the index, apply the reranker to this candidate set, and return the top 3 documents.

All retrieval models operate under a single-batch inference setup. Consequently, the end-to-end latency for queries \emph{without} reranking is identical across samples and is computed as:
\[
\text{latency per query} = 
\frac{\text{batch latency}}{\text{number of queries in the batch}}.
\]
This provides a consistent and fair latency comparison across all embedding-based retrieval baselines.

\section{Additional Analysis}
\label{apndx:analysis}

% \subsection{Queries Analysis}
% \label{apndx:queries_analysis}

% Figure~\ref{fig:features} reports which query features are statistically significant for each model, where a value of “1” denotes significance. The features themselves are defined in Table~\ref{tab:qualifiers}. Due to the relatively small number of queries, this analysis has certain limitations, and we exclude any feature that appears in fewer than 20\% of the queries. Each feature is treated as binary, indicating whether it occurs at least once within a given query.

% \begin{figure}[h]
% \centering
% \includegraphics[width=1.0\linewidth]{queries_analysis.png}
% \caption{Analysis features of the queries.}
% \label{fig:features}
% \end{figure}

% \subsection{Complexity}
% \label{sec:complx}

\subsection{Query Feature Analysis}
\label{apndx:queries_analysis}

\begin{figure}[h]
\centering
\includegraphics[width=1.0\linewidth]{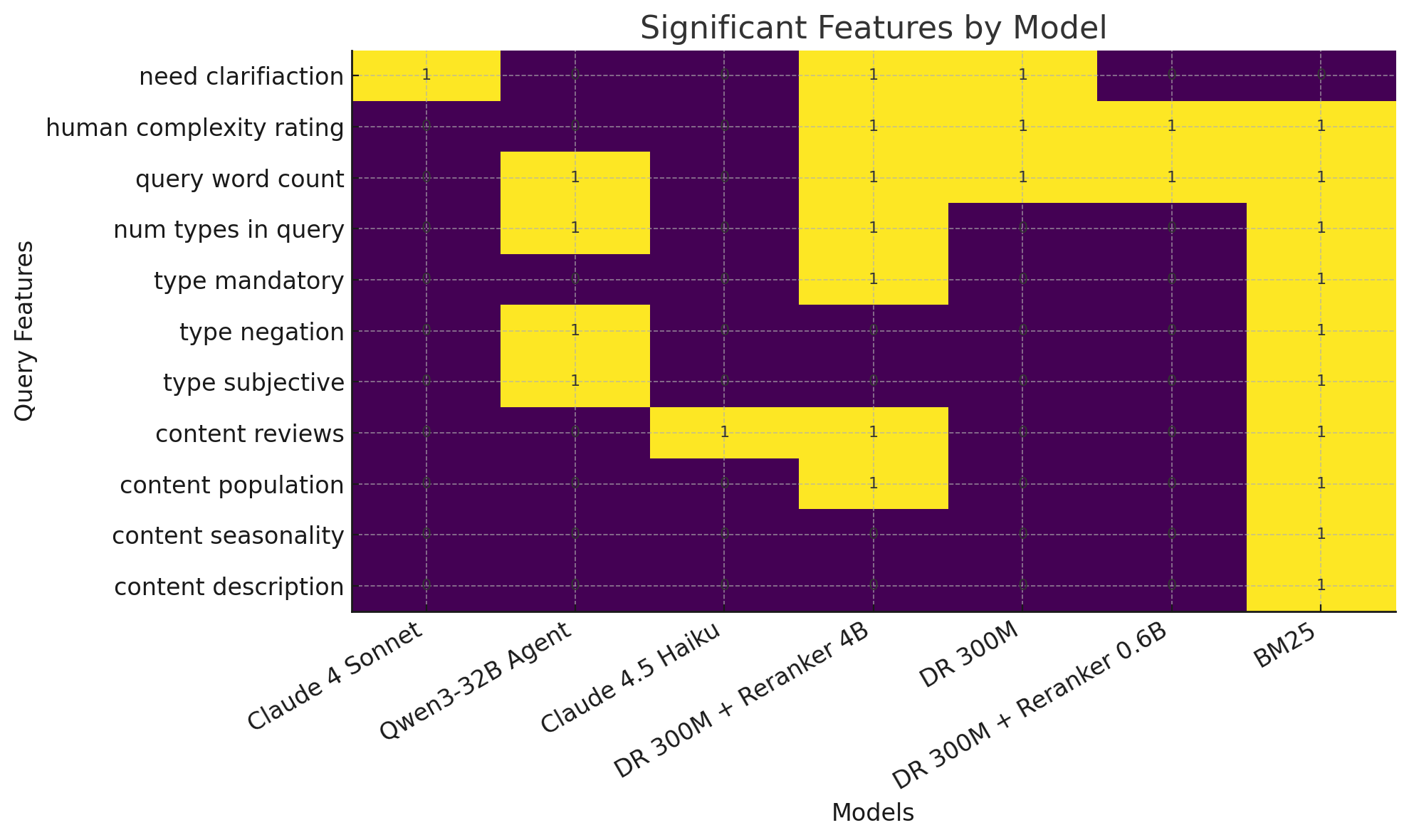}
\caption{Analysis of query features.}
\label{fig:features}
\end{figure}

Figure~\ref{fig:features} reports which query features are statistically significant for each model, where a value of “1” denotes significance. The features themselves are defined in Table~\ref{tab:qualifiers}. Due to the relatively small number of queries, this analysis has certain limitations, and we exclude any feature that appears in fewer than 20\% of the queries. Each feature is treated as binary, indicating whether it occurs at least once within a given query.

\subsection{Quality by Complexity}
We evaluate the agents within each complexity group to analyze how cost, token usage, latency, and accuracy vary as query difficulty increases. The results are presented in Table~\ref{tab:complexity_models_metrics}.

\begin{figure}[h]
\centering
\includegraphics[width=1\linewidth]{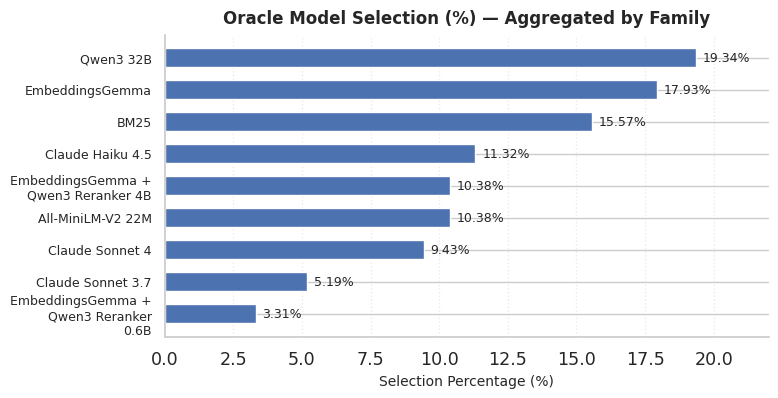}
\caption{Quality Oracle. Distribution of selected models when choosing, for each query, the cheapest model among those achieving the highest accuracy.}
\label{fig:oracle_2}
\end{figure}

\section{Evaluations}
\label{apndx:eval}

\begin{table*}[h!]
\centering
\small
\renewcommand{\arraystretch}{1.2}
\begin{tabular}{
    >{\raggedright\arraybackslash}p{0.05\textwidth}
    >{\raggedright\arraybackslash}p{0.10\textwidth}
    >{\raggedright\arraybackslash}p{0.25\textwidth}
    >{\raggedright\arraybackslash}p{0.40\textwidth}
}
\toprule
\textbf{Score} & \textbf{Label} & \textbf{Description} & \textbf{Criteria} \\
\midrule
\textbf{5} & Exact Match & The answer completely addresses all aspects of the query with specific, actionable hotel recommendations. &
$\bullet$ Addresses \emph{all} requirements (location, budget, amenities, group size, etc.)\\
& & & $\bullet$ Provides \emph{specific hotel names} and relevant details\\
& & & $\bullet$ Explains \emph{why} each recommendation fits \\
\midrule
\textbf{4} & Strong Match & Covers almost all requirements, with minor omissions or slight generalization. &
$\bullet$ Addresses \emph{most} requirements with relevant hotels\\
& & & $\bullet$ Missing minor detail (e.g., exact price or a less critical amenity) \\
\midrule
\textbf{3} & Partial Match & Covers some requirements but misses key aspects. &
$\bullet$ Addresses \emph{some} requirements\\
& & & $\bullet$ May give generic advice instead of specific hotels\\
& & & $\bullet$ Missing critical constraint(s) like budget, location, or amenities \\
\midrule
\textbf{2} & Weak Match & Provides tangentially relevant information but not directly aligned with query intent. &
$\bullet$ Hotel suggestions are only loosely related\\
& & & $\bullet$ Misses multiple key requirements\\
& & & $\bullet$ Possibly recommends wrong type of property or area \\
\midrule
\textbf{1} & Irrelevant & Fails to address the query requirements. &
$\bullet$ No relevant hotel recommendations\\
& & & $\bullet$ Wrong location/context\\
& & & $\bullet$ Ignores critical constraints \\
\bottomrule
\end{tabular}
\caption{Accuracy scoring rubric for hotel recommendation answers.}
\label{tab:hotel-rubric}
\end{table*}

\subsection{Human–LLM Agreement}
\label{app:agreement}

To evaluate the reliability of our automatic scoring pipeline, we measure the alignment between human judgments and LLM-based judgments. Specifically, we analyze different aggregation setups.

\begin{figure}[tbh]
\centering
\includegraphics[width=1.0\linewidth]{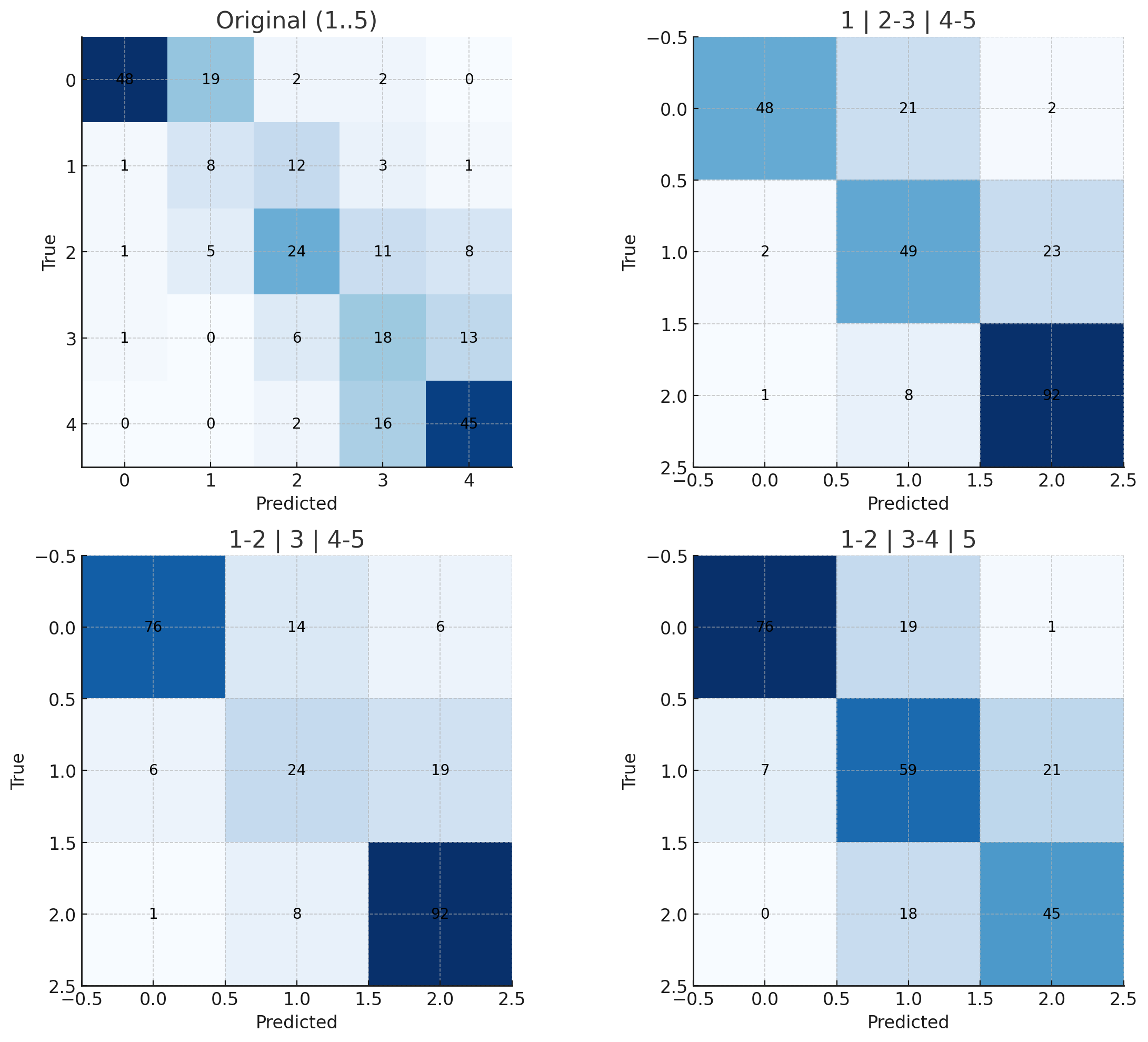}
\caption{Confusion matrices measuring alignment between human judgments and LLM-based scoring across different levels of label aggregation.}
\label{fig:confustion}
\end{figure}

Figure~\ref{fig:confustion} reports the resulting confusion matrices for each aggregation scheme. The matrices demonstrate strong alignment between human annotators and the LLM evaluator, with most disagreement concentrated in borderline or partially correct cases. This suggests that the LLM-based scoring mechanism is sufficiently reliable for large-scale evaluation while remaining sensitive to nuanced differences in answer quality. In total, we manually annotated \textbf{246 answers}, covering the complete spectrum of observed model behaviors.

Table~\ref{tab:agreement} reports the agreement between human annotators and the LLM-as-a-judge across multiple aggregation schemes of the 1–5 rating scale. The evaluation is based on \textbf{246 answers}, each independently rated by humans for different system baselines producing varying quality scores. We assess alignment using several complementary measures: (i) \textit{Exact Match}, capturing strict agreement; (ii) \textit{Cohen’s $\kappa$} with linear and quadratic weights, which account for partial disagreements and rating distance; and (iii) rank- and correlation-based measures, Spearman’s $\rho$, Kendall’s $\tau$, and Pearson’s $r$, to quantify ordinal and linear consistency. Additionally, non-parametric (\textit{Wilcoxon}) and parametric (\textit{paired $t$-test}) significance tests evaluate whether differences between distributions are statistically meaningful.

Across all aggregation schemes, correlations remain high ($\rho, r > 0.8$), and all tests indicate strong statistical significance ($p < 0.01$). This consistent alignment across diverse baselines and scoring distributions--demonstrating that the LLM-as-a-judge reliably mirrors human evaluation patterns, validating its use as a robust and scalable proxy for human judgment.

\begin{table*}[h!]
\centering
\small
\setlength{\tabcolsep}{5pt}
\begin{tabular}{lcccccccc}
\toprule
\textbf{Setting} & \textbf{Exact Match} & \textbf{$\kappa_{\text{linear}}$} & \textbf{$\kappa_{\text{quad}}$} & \textbf{$\rho$} & \textbf{$\tau$} & \textbf{$r$} & \textbf{Wilcoxon $p$} & \textbf{t-test $p$} \\
\midrule
Original (1–5)           & 0.5813 & 0.701 & 0.841 & 0.844** & 0.755** & 0.851** & ** & ** \\
Agg.: 1 \;|\; 2–3 \;|\; 4–5   & 0.7683 & 0.721 & 0.796 & 0.808** & 0.767** & 0.811** & ** & ** \\
Agg.: 1–2 \;|\; 3 \;|\; 4–5   & 0.7805 & 0.742 & 0.810 & 0.817** & 0.769** & 0.817** & ** & ** \\
Agg.: 1–2 \;|\; 3–4 \;|\; 5   & 0.7317 & 0.681 & 0.774 & 0.785** & 0.734** & 0.777** & * & * \\
\bottomrule
\end{tabular}
\caption{Agreement between human annotators and LLM-as-a-judge ratings.
** denotes $p < 0.01$; * denotes $p < 0.05$.}
\label{tab:agreement}
\end{table*}

\subsection{Factuality Evaluation Prompt}
\label{apndx:factuality}
For the Factuality Evaluation, we use a structured prompt that includes: 
(i)~a fixed evaluation header,  
(ii)~a placeholder describing the type of answer being evaluated,  
(iii)~the \textit{User Query} and the \textit{Model Answer},  
(iv)~the \textit{Clarification} (when applicable), and  
(v)~the complete \textit{Context} corresponding to all hotel documents cited by the model.  
This context consists of the full hotel descriptions and review texts associated with every citation the agent produces, as well as any snippets retrieved through web search when the agent invokes a web tool.
This setup ensures that the judge model evaluates factuality strictly based on verifiable evidence contained in the citations supplied by the agent.  
The full evaluation header used in the prompt is provided below.

\subsection{Accuracy Evaluation Prompt}
\label{apndx:acc}

For the Accuracy Evaluation, we use a structured prompt that includes:  
(i)~a fixed evaluation header,  
(ii)~a placeholder describing the type of answer being evaluated,  
(iii)~the \textit{User Query},  
(iv)~the \textit{Clarification} (when applicable), and  
(v)~the \textit{Model Answer}.  

This prompt focuses exclusively on how well the answer satisfies the user’s stated requirements, independent of factual grounding or citation quality.

The full evaluation header used in the prompt is provided below.

\onecolumn

\begin{lstlisting}[style=promptstyle]
You are a Factuality Judge for HOTEL RECOMMENDATIONS.
Your goal is to assess the factual accuracy of the ANSWER strictly
based on the provided hotel descriptions and reviews.
IGNORE any outside knowledge or assumptions ,  only consider information
verifiable from the given sources.

Task: Rate how FACTUALLY ACCURATE the ANSWER is on a 1 to 5 scale:
1 = Completely inaccurate: contains mostly false or unsupported statements.
2 = Poor factualit: some facts are correct, but most claims lack evidence or 
contradict the sources.
3 = Partially factual: roughly half the claims are supported, others are vague or unverified.
4 = Mostly factual: nearly all claims align with the sources, with only minor 
inaccuracies or omissions.
5 = Fully factual: every factual statement is accurate and directly supported by a cited source.

When evaluating, consider:
- Does each factual statement about the hotel (e.g., location, amenities, ratings, 
accessibility, services)
  have explicit evidence from the provided descriptions or reviews?
- Are there any hallucinated details or claims not grounded in the sources?
- Are sources cited clearly and correctly linked to each factual statement?
- Is the information consistent with the evidence, without contradictions or exaggerations?
- IMPORTANT: If any factual statement lacks an explicit source, deduct points proportionally.

Output format: Return ONLY a valid JSON object with two fields:
- score: an integer from 1 to 5
- explanation: a concise justification mentioning which parts are well-supported and which are not.

Example:
{
  "score": 4,
  "explanation": "Most details (location, breakfast, and accessibility) are supported by the
  descriptions, but the mention of a rooftop bar lacks evidence."
}

Do not include any text outside the JSON object.
\end{lstlisting}

\begin{lstlisting}[style=promptstyle]
You are a Relevance Judge for HOTEL RECOMMENDATIONS.
Evaluate ONLY using the provided hotel descriptions and reviews 
(ignore any outside knowledge).

Task: Rate how well the ANSWER addresses the USER QUERY on a 1 to 5 scale:
1 = Not relevant at all: completely misses the user's needs.
2 = Slightly relevant: mentions minor aspects but not the core requirements.
3 = Moderately relevant: covers some key points but ignores important requirements.
4 = Very relevant: satisfies most requirements with only minor omissions.
5 = Perfectly relevant: fully addresses all requirements with appropriate detail.

When evaluating, consider:
- Does the answer directly address the specific hotel requirements 
  (location, budget, amenities, travel dates, party size)?
- Are concrete hotel recommendations provided (hotel names + pertinent details), 
  rather than generic or high-level advice?
- Is the reasoning clear, structured, and grounded in the provided descriptions/reviews?
- Are trade-offs or limitations explained when relevant?
- IMPORTANT: If the query requires recommending hotels and the answer does NOT 
  provide any concrete hotel recommendation, score = 1.

Output format: Return ONLY a valid JSON object with two fields:
- score: an integer from 1 to 5
- explanation: a brief justification for the chosen score.

Example:
{
  "score": 4,
  "explanation": "The answer addresses most user requirements and provides hotel names,
  but it lacks detail about budget constraints."
}

Do not include any text outside the JSON object.
\end{lstlisting}

\twocolumn

\end{document}